\RequirePackage{fix-cm}  % to remove annoying warnings
\documentclass[aps,prl,showpacs,amsmath,amssymb,amsfonts,superscriptaddress,twocolumn,lengthcheck]{revtex4-1}
\usepackage{graphicx}
\usepackage{subfigure}
\usepackage{amsthm}
\usepackage{verbatim}
\usepackage{dcolumn}% Align table columns on decimal point
\usepackage{bm}% bold math
\usepackage{epsf}
\usepackage{color}
\usepackage[colorlinks=true,citecolor=blue,linkcolor=blue,urlcolor=blue]{hyperref}%
\newcommand{\bra}[1]{\left\langle #1\right|}
\newcommand{\ket}[1]{\left|#1\right\rangle}
\newcommand{\braket}[2]{\left\langle #1|#2\right\rangle}

\newcommand{\id}{\mathbb{I}}

\newcommand{\bla}{bla\\bla\\bla\\bla\\bla}

\begin{document}
\title{Repeatability of measurements: Equivalence of hermitian and non-hermitian observables}
\author{Bart\l{}omiej Gardas}
\affiliation{Theoretical Division, Los Alamos National Laboratory, Los Alamos, NM 87545, USA}
\affiliation{Institute of Physics, University of Silesia, 40-007 Katowice, Poland}
\author{Sebastian Deffner}
\author{Avadh Saxena}
\affiliation{Theoretical Division, Los Alamos National Laboratory, Los Alamos, NM 87545, USA}
\affiliation{Center for Nonlinear Studies, Los Alamos National Laboratory, Los Alamos, NM 87545, USA}
\date{\today}

\begin{abstract}	
A non-commuting measurement transfers, via the apparatus, information encoded in a system's state to the external ``observer''.
Classical measurements determine properties of physical objects. In the quantum realm, the very same notion restricts the recording
process to orthogonal states as only those are distinguishable by measurements. Therefore, even a possibility to describe 
physical reality by means of non-hermitian operators should \emph{volens nolens} be excluded as their eigenstates are not orthogonal.
Here, we show that non-hermitian operators with real spectrum can be treated within the standard framework of quantum mechanics. 
Furthermore, we propose a quantum canonical transformation that maps hermitian systems onto non-hermitian ones. Similar to classical 
inertial forces this transformation is accompanied by an energetic cost pinning the system on the unitary path.
\end{abstract}

 % Quantum mechanics, Foundations of quantum mechanics; measurement theory, Formalism
\maketitle
 
\paragraph{Introduction.}
The no-cloning theorem states that unknown quantum states cannot be copied~\cite{nocloning}, since no measurement can distinguish
arbitrary states with certainty. Similarly, the unitary transfer of information from a quantum system to the measuring device -- apparatus -- 
cannot distinguish between non-orthogonal states~\cite{woj_rep_1,woj_rep_2}. In contrast, all physical properties of classical systems can 
be determined with arbitrary precision as the recording process does not perturb the system. To put it differently, classical measurements 
do not involve any back-action.

Those profound facts rely on a rather natural assumption regarding the physical world -- the repeatability of measurements. The latter requires
that consecutive measurements should result in the same outcome. Consequently, the demand for all physical observables to be hermitian seems to be 
justified from the ``first principles''~\cite{woj_rep_2}. Therefore, even a possibility to represent observables using non-hermitian operators
should \emph{volens nolens} be excluded as their eigenstates are non-orthogonal~\cite{Moiseyev2011}.
 
Nevertheless, a variety of experimental findings can be explained by means of non-hermitian operators. For instance, a spontaneous symmetry 
breaking observed in Ref.~\cite{optic,optic_2} has been linked to $\mathcal{P}\mathcal{T}$-symmetry, a condition weaker than hermiticity~\cite{brody}. 
Here $\mathcal{P}$ and $\mathcal{T}$ denote the parity and time reflection, respectively, and $\mathcal{P}\mathcal{T}$-symmetry guaranties that
$[\mathcal{P}\mathcal{T},H]=0$, where $H$ is the system's Hamiltonian. Additionally, exceptional eigenenergies of complex value have also been 
measured~\cite{gao}. Recent years have witnessed great theoretical progress towards the understanding of as such non-hermitian 
systems~\cite{Moiseyev2011}. It has been shown, for example, that conventional quantum mechanics can be extended to the complex 
domain~\cite{bender_1}. Interesting examples are optical systems with complex index of refraction~\cite{Berry2011}, tilted optical lattices 
with defects~\cite{Wunner16} or systems undergoing topological transitions~\cite{Levitov09,Zeuner15}. The latter can serve as realizations of $\mathcal{P}\mathcal{T}$-symmetry in Bose-Einstein condensates~\cite{Stringari99}. Also, many breakthroughs in thermodynamics and statistical
physics have been reported for non-hermitian systems. The Jarzynski equality~\cite{pt} or the Carnot bound~\cite{pseudo-thermo} may serve 
as good examples~\cite{brody,bender03}. % Later: cite Song15

It is only natural to ask whether such theories are fundamental or provide only an \emph{effective} (\emph{e.g.} open systems with balanced loss
and gain~\cite{optic}) description of nature~\cite{Meng15}. In this Letter we show that the requirement to be able to repeat measurements does not
exclude \emph{all} non-hermitian ``observables" from the description of physical reality. We will prove that the non-hermitian observables with real 
spectrum are as physical as their hermitian counterparts. In fact, a formal correspondence between the two classes can be established by means of a 
quantum canonical transformation~\cite{Anderson94}. To put it differently, non-hermitian operators provide a convenient way of representing quantum 
systems in a physically equivalent way~\cite{ali07,aliPRL07}. This situation is completely analogous to classical mechanics where classical canonical 
transformations are used to simplify Hamilton's equations of motion~\cite{Goldstein02}.
 
\paragraph{Repeatability of quantum measurements.}

Let $H$ be a non-hermitian observable, \emph{i.e.} $H^{\dagger}\not=H$. Without loss of generality we assume that $H$ is the Hamiltonain of
a quantum system. To be physically relevant $H$ needs to be \emph{at least} diagonalizable. This requirement assures the existence of an orthonormal 
basis, $\{\ket{E_n}\}$ with corresponding eigenenergies, $E_n$, that can be measured. For the sake of simplicity, we further assume that the energy 
spectrum is discrete and non-degenerate. Therefore, we can write~\cite{strang06}

\begin{equation} 
	\label{diag}
	V^{-1} H V = \sum_{n}E_n\ket{E_n} \bra{E_n},
	\quad
	E_n\in\mathbb{C},
\end{equation}
where $\braket{E_n}{E_m}=\delta_{nm}$.
Generally, $H$ is non-hermitian, and thus the similarity transformation $V$ is not unitary, \emph{i.e}, $V^{\dagger} \not =V^{-1}$.
Let us rewrite Eq.~(\ref{diag}) as

\begin{equation}
	% H = \sum_{n}E_n(V\ket{E_n}) (\bra{E_n}V^{-1}) \equiv \sum_{n}E_n\ket{\psi_n}\bra{\phi_n}.
	  H = \sum_{n}E_n\ket{\psi_n}\bra{\phi_n},
\end{equation}
where $\ket{\psi_n}:=V\ket{E_n}$ and $\bra{\phi_n}:=\bra{E_n}V^{-1}$. By construction, these states form a \emph{biorthonormal} 
basis~\cite{brody,avadh08}, that is, $\braket{\phi_n}{\psi_m} = \delta_{nm}$ and $\id = \sum_{n}\ket{\psi_n}\bra{\phi_n}$.
Moreover, the eigenvalue problem for $H$ can be stated as

\begin{equation}
	H\ket{\psi_n} =E_n\ket{\psi_n}, \quad
	%
	%H^{\dagger}\ket{\phi_n} =E_n^*\ket{\phi_n}. 
	\bra{\phi_n}H =E_n\bra{\phi_n}.
\end{equation}
As a result, the left $\bra{\phi_n}$ corresponds to the right $\ket{\psi_n}$ in Dirac notation~\cite{Dirac1939}.
Hence, the way the probability is assigned to a physical process has to be revisited. For instance, consider a system that is
prepared in one of its energy eigenstates, say $\ket{\psi_n}$, and then immediately perturbed by a map $U$ (\emph{e.g.} time 
evolution or a measurement). Then the probability to measure energy $E_m$ reads 

\begin{equation}
	p_{nm} = p(\ket{\psi_n}\rightarrow U \rightarrow \ket{\psi_m}) = \big|\bra{\phi_m}U\ket{\psi_n}\big |^2.
\end{equation}%
This formula provides a \emph{natural} generalization of the ``standard'' recipe, $|\bra{\psi_m}U\ket{\psi_n}|^2$, for calculating
probabilities in hermitian quantum mechanics~\citep{Dirac1939}.
 
It is important to realize that non-hermitian operators such as in Eq.~(\ref{diag}) \emph{can} generate unitary dynamics. This is possible 
\emph{if and only if} the energy spectrum is real. Therefore, \emph{a priori}, non-hermitian operators do not necessarily violate the 
conservation of probability. Note, not only every superposition of states $\ket{\psi_n}$ is allowed but also an arbitrary state $\ket{\psi_0}$
can be expressed in such a manner, $\ket{\psi_0} = \sum_n c_n \ket{\psi_n}$, where $\sum_n|c_n|^2=1$ and $|c_n|^2$ is the probability for the 
system to be found in its eigenstate $\ket{\psi_n}$. The corresponding $\bra{\phi_0}$ is given by 

\begin{equation}
\label{bra}
\bra{\phi_0} = \sum_n c_n^* \bra{\phi_n}.
\end{equation} 
The initial state $\ket{\psi_0}$ evolves under the Schr\"odinger equation, $i\hbar\partial_t\ket{\psi_t}=H\ket{\psi_t}$. Therefore, the 
solution reads $\ket{\psi_t} = \sum_n e^{-iE_nt/\hbar} c_n \ket{\psi_n}$, and, as a consequence of Eq.~(\ref{bra}), the corresponding 
$\bra{\phi_t}$ evolves according to
\begin{equation}
\bra{\phi_t} = \sum_n e^{+iE_n^*t/\hbar} c_n^* \bra{\phi_n}.
\end{equation}
Thus, \emph{if and only if} all eigenvalues $E_n$ are real then the system's dynamics is unitary and therefore

\begin{equation}
	\braket{\phi_t}{\psi_t} = \sum_n  |c_n|^2 e^{i\left(E_n-E_n^*\right)t/\hbar} = \braket{\phi_0}{\psi_0},
\end{equation}
proving that the probability is indeed conserved.

We have demonstrated that a fully consistent quantum theory can be built with non-hermitian operators. We have imposed that ``observables'' 
are diagonalizable, which assures that the spectrum can be measured. Its reality, on the other hand, yields unitary dynamics and thus the 
conservation of probability. 

Let $\ket{A_0}$ be a ``ready to measure'' initial state of the apparatus $\mathcal{A}$. Further, by $\ket{\psi_n}$, $\ket{\psi_m}$ we denote
distinct ($n\not=m$) eigenstates of $H$. Also, we assume that $E_n^*=E_n$ and without loss of generality, we choose $\mathcal{A}$ to be a 
hermitian system. The repeatability of measurements guarantees that every unitary transfer of information from system $\mathcal{S}$ to $\mathcal{A}$
leaves states $\ket{\psi_n}$ and $\ket{\psi_m}$ undisturbed. It follows that

\begin{figure}
	\includegraphics[width=0.48\textwidth]{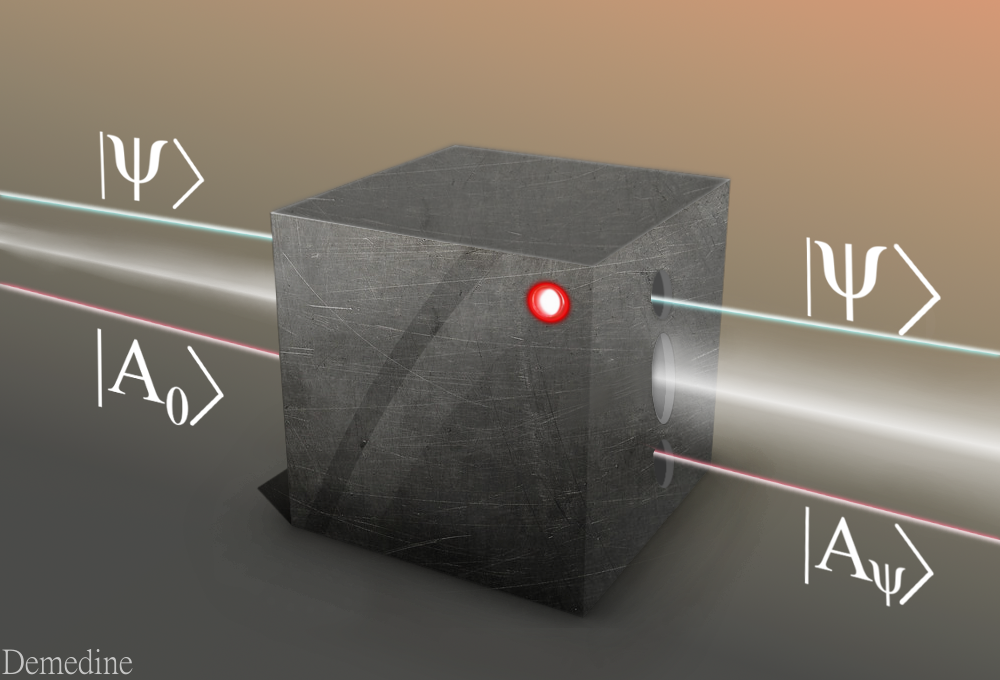}
	\caption{\label{fig:apparatus} (Color online)
		Schematic representation of the unitary transfer of information from a quantum system $\mathcal{S}$ 
		to the measuring device -- apparatus $\mathcal{A}$. Information encoded in a quantum state $\ket{\psi}$
		is being recorded by $\mathcal{A}$ and is written down into its state $\ket{A_{\psi}}$. The recording process
		$\ket{A_0} \rightarrow \ket{A_{\psi}}$ is unitary and does not influence the information carried by $\ket{\psi}$.
	}
\end{figure}

\begin{equation}
\label{measure}
\mathcal{U}: \ket{\psi_k}\ket{A_0} \rightarrow \ket{\psi_k}\ket{A_k}
\quad
\text{for}
\quad k=n,m,
\end{equation}
where $\mathcal{U}$ is a unitary map (\emph{e.g.} $\mathcal{U}\mathcal{U}^{\dagger}=\id$) modeling the recording process. As illustrated 
in Fig.~\ref{fig:apparatus}, after the transfer has been completed, new states $\ket{A_m}$ and $\ket{A_n}$ of the apparatus $\mathcal{A}$
encode the information about the system's eigenstates $\ket{\psi_n}$ and $\ket{\psi_m}$. The measurement preserves the norm on the Hilbert
space $\mathcal{S}\otimes\mathcal{A}$ as well. Hence
\begin{equation}
\label{right}
\braket{\phi_m}{\psi_n}\left( 1 - \braket{A_m}{A_n}\right) = 0.
\end{equation}
As a result, the apparatus states $\ket{A_m}$ and $\ket{A_n}$ can differ (indicating some information being stored) only when $\braket{\phi_m}
{\psi_n}=0$. To put it differently, extracting information that distinguishes between two measured states is possible only when the transition 
probability between state $\ket{\psi_n}$ and $\ket{\psi_m}$ vanishes.

The above analysis demonstrates that the reality of the spectrum, which guarantees unitarity, rather that hermiticity is \emph{necessary} 
to acquire information. We stress that $|\braket{\psi_m}{\psi_n}|^2$ is \emph{not} the transition probability between states $\ket{\psi_n}$
and $\ket{\psi_m}$. That is given by $|\braket{\phi_m}{\psi_n}|^2$ and the two coincide only when $H$ is hermitian as only then the left
and right eigenstates are the same. 

Adopting the similar formula that was derived for hermitian systems (see \emph{e.g.} Eq.~$2b$ in~\cite{woj_rep_1}) one could write
$\braket{\psi_m}{\psi_n}\left( 1 - \braket{A_m}{A_n}\right) = 0$. As a result, one would have to conclude that $\braket{A_m}{A_n}=1$ showing 
the apparatus cannot tell the measured states apart. However, as we have shown, since $H$ is non-hermitian Eq.~(\ref{right}) applies, which 
allows non-orthogonal states to be measured.
\paragraph{Relation to hermitian systems.}

Thus far we have seen that the proper identification of probabilities of the measurement outcomes allows to include non-hermitian
operators into the usual framework of quantum mechanics rather naturally. This is done, in a physically consistent way, by properly
accounting for the fact that non-hermitian observables have different left and right eigenstates. The probability 
$p_n=|\braket{\phi_n}{\psi}|^2$ to find the system in its eigenstate $\ket{\psi_n}$ can also be rewritten, using this state \emph{explicitly},~as
\begin{equation}
\label{metric}
p_n = |\left(\bra{\psi_n} g\right)\ket{\psi}|^2,
\quad
g = \sum_{m}\ket{\phi_m}\bra{\phi_m}.
\end{equation}
Thus, the Dirac correspondence between bra and ket vectors can now be understood as $\bra{\psi}g\leftrightarrow\ket{\psi}$. Above, $g$
is a positive-definite, invertible linear operator -- that is a metric~\cite{ap}. Indeed, we have

\begin{equation}
\label{pos}
\bra{\psi}g\ket{\psi} = \sum_{m}|\braket{\phi_m}{\psi}|^2, % > 0,
\,
g^{-1} = \sum_{n}\ket{\psi_n}\bra{\psi_n},
\end{equation}
where the first equality expresses positivity and the second expression provides an explicit formula for the inverse map in terms of 
eigenstates $\ket{\psi_n}$. As a result, assigning probabilities when measuring non-hermitian ``observables'' defines a new inner product,
namely $(\psi,g\phi)$. Moreover, a simple calculation shows that $H^{\dagger}g=gH$ and therefore $(H\psi,g\phi) = (\psi,gH\phi)$ for all 
states $\ket{\psi}$, $\ket{\phi}$. This fact can also be interpreted as: $H$ is hermitian with respect to this new inner product.

In conclusion, from the viewpoint of a measurement, there is no \emph{physical} difference between non-hermitian operators with
real spectrum and hermitian observables. Therefore, one should be able to represent quantum systems either way depending on the
situation. Of course, in complete analogy to classical physics the goal is to find the simplest possible Hamiltonian. 
One can establish a correspondence between $H$ and its hermitian counterpart $K$ in the following way~\cite{Ali05}

\begin{equation}
\label{HK}
	\begin{split}
	 K &=  g^{1/2} H g^{-1/2} = e^{G/2} H e^{-G/2}   \\
	   &= H + \frac{1}{2} \left[G,H\right] + \frac{1}{2!2^2} \left[G,\left[G,H\right]\right] + \cdots
	\end{split}   
\end{equation}
where $G:=\ln(g)$. In the second line we have used the Baker-Campbell-Hausdorff like formula~\cite{Rossmann06}. Note, since the metric
$g$ is positive definite its logarithm and square root exist and moreover, both of these quantities are hermitian operators~\cite{reed78}.
\footnote{Equation.~(\ref{HK}) was introduced in~\cite{Ali05} as a similarity map between Hilbert spaces. Physical significance was missed}
Although this infinite series does not truncate after a finite number of terms in general, there are physically relevant examples where it
does (see \emph{Example 2} below). We stress that Eq.~(\ref{HK}) can be used to transform an arbitrary observable $O$ between hermitian and 
non-hermitian representations.

First of the above equations~(\ref{HK}) shows that $K$ is indeed hermitian, whereas the second line demonstrates an interesting 
feature of physical reality. Namely, a quantum system can be represented \emph{equivalently} either by a hermitian operator or a non-hermitian
one with real spectrum. Although there is no essential physical difference between the two representations, their mathematical structures are
quite different. It follows directly from Eq.~(\ref{HK}) that a complicated hermitian system may have a very simple non-hermitian representation
and \emph{vice versa}. Transformation~(\ref{HK}) plays an analogous role to the canonical transformation well established in classical 
mechanics~\cite{Goldstein02}. Note, this transformation cannot be unitary as it changes the hermiticity of an operator. However, it 
preserves the canonical commutation relation and as such belongs to a class of quantum canonical transformations~\cite{Anderson94,Lee95}.

In classical mechanics, Newtonian equations of motion have to be modified in non-standard, time-dependent, frames of reference~\cite{Goldstein02}. 
As a result, one observes so-called inertial forces. Typical examples include Coriolis or centrifugal forces that are present \emph{only} in 
rotating frames of reference. Therefore, there are experimentally accessible consequences of using such non-inertial coordinates. One of the most 
famous examples is the Foucault pendulum whose motion (precession) directly reflects on the Earth's rotation around its own axis~\cite{F14,oprea95}. 

Interestingly, if the non-hermitian Hamiltonian is time-dependent then the corresponding Schr\"odinger equation also has to be modified 
to preserve unitarity~\cite{Znojil,gong,Znojil08},

\begin{equation}
\label{mse}
i\hbar\partial_t \ket{\psi_t} = \left(H_t +F_t\right)\ket{\psi_t},
\
F_t=-\frac{i\hbar}{2} \Lambda_t^{-1} \partial_t \Lambda_t. 
\end{equation}
Above, $\Lambda_t$ is a time-dependent metric that does not necessarily coincide with $g_t$~\cite{gong}. More importantly, this metric is not 
unique. However, it can be chosen so that the corresponding hermitian Hamiltonian $K_t$ in Eq.~(\ref{HK}) (i) is the generator of dynamics and
(ii) has exactly the same spectrum as $H_t$. 

Therefore, the dynamics in these two representations differ considerably. Nevertheless, replacing $\partial_t$ with a covariant derivative 
$D_t:=\partial_t+\Lambda_t^{-1} \partial_t \Lambda_t/2$~\cite{Luc01} the Schr\"odinger Eq.~(\ref{mse}) can be put into its standard form,
\emph{i.e.}, with $H_t$ being the generator, $i\hbar D_t \ket{\psi_t} = H_t\ket{\psi_t}$. However, one can also think of the extra energetic 
contribution $ \sim \Lambda_t^{-1} \partial_t \Lambda_t/2$ as being a manifestation of a force of inertia keeping a quantum system along the
unitary path during its evolution (see \emph{Example~3} below). 

It is worth mentioning that the existence of $F_t$ in non-hermitian representations has already been noticed yet disregarded as unphysical
(see for example~\cite{ali07} and comments that followed). It was treated rather as a mathematical necessity not having much to do with 
physical reality. We will now illustrate the novel concepts with several analytically tractebel example.

\paragraph{\label{ex:1} Example 1: Paradigmatic  $\mathcal{P}\mathcal{T}$ - symmetric system.}
As a first example consider a harmonic oscillator with a non-hermitian perturbation, for instance~\cite{Bender04} % Erratum:,*Bender05},

\begin{equation}
\label{ho3}
H = \frac{p^2}{2m} + \frac{1}{2}m\omega^2 x^2 + i \epsilon x^3 \equiv H_0 + \epsilon H_1,
\end{equation}
where $H_0$ corresponds to the unperturbed harmonic oscillator and $H_1$ is an anharmonic perturbation. Parameters $m$ and $\omega$ 
correspond to the system's mass and the size of the harmonic trap, respectively. Here $\epsilon$ is a small perturbation. The momentum $p$
and position $x$ operators obey the standard canonical commuatation relation, $[x,p]=i\hbar$. This model has been extensively investigated 
in literature~\cite{Croke12}. Numerical studies have confirmed the reality of its spectrum for all real $\epsilon$. Using perturbation 
theory one can establish that $K=H_0+V(x,p)$ where the momentum dependent potential $V(x,p)$ up to ${\cal O}(\epsilon^3)$ reads~\cite{Ali05}

\begin{equation}
\label{Vxp} 
    V=\frac{1}{m\omega^4}\left(\{x^2,p^2\}+p\,x^2p+\frac{3m\omega^2}{2}\,x^4\right)  \epsilon^2,
\end{equation} 
where $\{A,B\}:=AB+BA$ is the anticommutator between $A$ and $B$. Observe that this very complicated, momentum dependent potential
can \emph{effectively} be replaced by a simple non-hermitian term $V \sim ix^3$, which only depends on the position $x$~\cite{cho}.

\paragraph{\label{ex:2} Example 2: Localization in condensed matter physics.}
Another interesting example is a general one dimensional quantum system whose Hamiltonian reads $K=p^2/2m + V(x)$, where $V(x)$ is
an arbitrary potential. This standard textbook hermitian model can be turned into a very powerful non-hermitian system that can
explain localization effects in solid state physics~\cite{Longhi10,Longhi13}. Indeed, we have

\begin{equation}
\label{nherm}
		H   = e^{-\eta x} K e^{\eta x}  
		    %= \frac{1}{2m}\left(p -\eta\left[x,p\right]+ \cdots\right)^2  + V(x)\\
	  	    = \frac{(p-i\hbar\eta)^2}{2m} + V(x), 	    
\end{equation} 
where the real parameter $\eta$ expresses an external magnetic field~\cite{Hatano96}. Note that the metric $g=e^{2\eta x}$ in this case
can be calculated explicitly. Furthermore, it depends on the external control parameter -- the magnetic field~\cite{brody15}. Also, 
since the commutator $[x,p]=i\hbar$ is a complex number the infinite series in Eq.~(\ref{HK}) truncates after only two terms.

\begin{figure}[t]
	\includegraphics[width=0.48\textwidth]{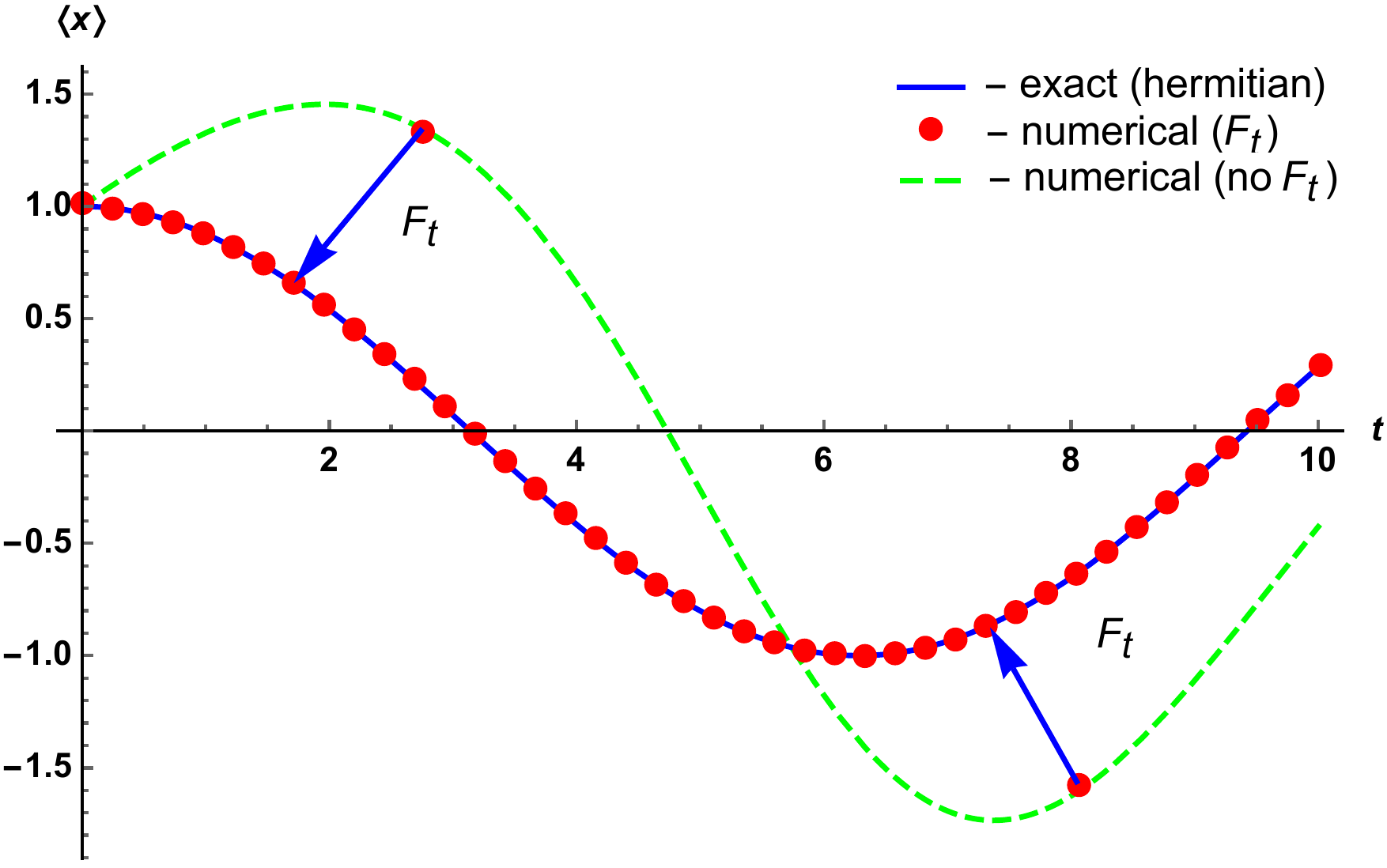}
	\caption{\label{fig:F} (Color online) The average position $\langle x\rangle$ of a quantum particle  as a function of time $t$. 
		Blue solid curve is the exact solution to the Schr\"odinger equation obtained in the hermitian representation. 
		Red points represent a numerical solution to the same problem computed in the non-hermitian (time-dependent)
		representation. The two paths coincide only if the inertial force is accounted for [see Eq.~(\ref{Ft})]. 
		Finally, green dashed line depicts a ``naive'' solution obtained without taking into account this contribution. 
		Parameters are $\hbar=m=1.0$, $\omega=0.5$ and $\eta_t=t/\tau$, where $\tau=10$. The initial state is $\ket{\psi_0}=(\ket{0}+\ket{1})/\sqrt{2}$, where $\ket{0}$ is the ground state of $K=H_0$ and $\ket{1}=a^{\dagger}\ket{0}$.
	}
\end{figure}

\paragraph{Example 3: Time-dependent metric and force of inertia.}\label{ex:3}
Finally, assume that the metric $g$ from the previous example depends explicitly on time (\emph{e.g.} the magnetic field $\eta_t$ is
time-dependent). We further choose $V(x)$ to be a harmonic trap, $V(x)=m\omega^2 x^2/2$, where $\omega$ is its frequency. Then the 
Hamiltonian in Eq.~(\ref{nherm}), now time-dependent, can be written using the second quantization~\cite{Girardeau71} as 

\begin{equation}
\label{nho}
H_t = \hbar\omega \left[\left(a-\eta_t\alpha\right)^{\dagger} \left(a+\eta_t\alpha\right) + \frac{1}{2}\right],
\end{equation}
where $\alpha=\sqrt{\hbar/2m\omega}$ and $a$, $a^{\dagger}$ are annihilation and creation operators, respectively.
%~\footnote{In the second quantization we have $[a,a^{\dagger}]=\id$.}. 
As explained above, to preserve unitarity the evolution generator $H_t$ in the non-hermitian 
representation has to be modified accordingly. By setting $\Lambda_t=g_t$~\cite{pseudo-thermo} in Eq.~(\ref{mse}) we have 

\begin{equation}
\label{Ft}
H_t\rightarrow H_t + F_t, \quad F_t = -i\dot{\eta_t}\alpha\left(a+a^{\dagger}\right).
\end{equation}
To analyze the evolution of this system we turn to numerical simulations. We further assume that $\eta_t$ changes on a time scale $\tau$ 
linearly, that is, $\eta_t=t/\tau$ for $0\leq t\leq\tau$. The initial state is given by $\ket{\psi_0}=(\ket{0}+\ket{1})/\sqrt{2}$, where 
$\ket{0}$ is the ground state of $K=H_0$ and $\ket{1}=a^{\dagger}\ket{0}$. 

Figure~\ref{fig:F} shows the average position $\langle x\rangle$ of a quantum particle as a function of time $t$ computed both in the hermitian 
(blue solid line) as well as the non-hermitian (red points) representation. According to the Ehrenfest theorem, $\langle x\rangle$ corresponds 
to the classical trajectory in a sense that it obeys Newton's equations of motion~\cite{Ehrenfest27}. As we can see, only when a proper energetic 
contribution $F_t$ is accounted for the two paths coincide. The Dashed green line, on the other hand, depicts a non-unitary path resulting from
not taking into account this contribution. As expected, $F_t$ does not have any influence on the system's dynamics in non-accelerating frames of
reference where $\dot{\eta}_t=0$.

\paragraph{Summary.}
    
The very question whether physical observables should be hermitian or not reflects on a long lasting debate regarding physical reality.
Nowadays, this issue is no longer only of academic interest as leading groups are beginning to investigate it experimentally 
(\emph{e.g.} ~\cite{optic,gao}). 
In this Letter we have revisited this problem showing that the repeatability of measurements does not exclude non-hermitian operators from 
the usual framework of quantum mechanics. We have argued that operators which admit real spectrum are canonicaly equivalent to hermitian ones.
As a result, all fundamental notions (\emph{e.g.} repeatability of a measurement, no cloning theorem, etc.) that have been associated with the 
unitarity apply to all non-hermitian systems with real spectrum as well.

The question which of these two representations is more adequate to describe a quantum system depends on the problem under investigation. 
It may be more natural to use a non-hermitain frame of reference. However in that case, as a result of using a non-standard representation,
the resulting Schr\"odinger equation has to be modified accordingly [see Eq.~(\ref{mse})]. There is an extra energetic contribution that has
to be accounted for to preserve unitarity. We have associated this energetic cost $F_t$ with a inertial force that keeps a quantum system on 
the unitary path during its evolution (see \emph{Example 3}). As it is in classical mechanics, $F_t$ vanishes for all non-accelerating frames
of reference, \emph{i.e.} with $\dot{g}_t=0$.
  
We should stress here that not all non-hermitian systems have real spectrum. Those whose eigenenergies (at least some of them) are complex 
were explicitly excluded from our considerations. Such systems are \emph{open}~\cite{breuer02}. During their evolution they lose or gain energy 
and information in a way that cannot be balanced~\cite{Moiseyev2011}. Therefore, a unitary map is not sufficient to capture their dynamics anymore. 
Interesting examples can be found \emph{e.g.} in~\cite{gao,Berry2011}.
   
\begin{acknowledgments}
\emph{Acknowledgments}. 
It is our pleasure to thank Wojciech H. Zurek and Rolando Somma for stimulating discussions. We gratefully acknowledge Marta
Paczy{\'n}ska who designed and prepared Fig.~\ref{fig:apparatus}. This work was supported by the Polish Ministry of Science
and Higher Education under project Mobility Plus 1060/MOB/2013/0 (B.G.); S.D. acknowledges financial support from the U.S. 
Department of Energy through a LANL Director's Funded Fellowship.
\end{acknowledgments}
%
%\bibliography{rep}

\end{document}